\documentclass{phyeauth}

\usepackage{graphicx}
\usepackage{amsmath,amssymb,mathrsfs} 
\usepackage{bm}
\usepackage{bbm}
\usepackage{psfrag}
\usepackage{latexsym}
\usepackage{subfigure}     
\usepackage{stmaryrd}
\newcommand{\n}{\nonumber} 
\newcommand{\refsec}[1]{Sect.~\ref{#1}}
\newcommand{\reffig}[1]{Fig.~\ref{#1}}

\newlength{\AbstandAutor}
\newlength{\AbstandZitat}
\renewcommand{\thefootnote}{\fnsymbol{footnote}}

\newcommand{\name}[1]{#1}
\newcommand{\qmarks}[1]{``#1"}

\newcommand{\hilbertspace}{\name{Hilbert} space}
\newcommand{\sbild}{\name{Schr\"odinger} picture}
\newcommand{\ibild}{interaction picture}
\newcommand{\sgleichung}{\name{Schr\"odinger} equation} 

\newcommand{\hamiltonian}{Hamiltonian}
\newcommand{\einsop}{$\1$-operator}

\newcommand{\FgoldenRule}{\name{Fermi}'s Golden Rule}

\newcommand{\Gaussian}{\name{Gauss}ian}

\newcommand{\hsav}{\hilbertspace\ average}

\newcommand{\fewlevel}{few-level}

\newcommand{\timedependent}{time-de\-pend\-ent}

\newcommand{\stepwise}{step-wise}
\newcommand{\timestep}{time step}

\newcommand{\offdiagonal}{off-diagonal}

\newcommand{\precondition}{pre-condition} 
\newcommand{\ie}{i.e.}
\newcommand{\eg}{e.g.}

\newcommand{\setup}{set-up}

\newcommand{\energyconservation}{energy conservation}

\mathchardef\ordinarycolon\mathcode`\:
\mathcode`\:=\string"8000
\begingroup \catcode`\:=\active
  \gdef:{\mathrel{\mathop\ordinarycolon}}
\endgroup
\newcommand{\Funktion}[2]{#1\kern-0.2em\left(#2\right)}
\newcommand{\FunktionNeu}[2]{#1\kern-0.1em\big(#2\big)}

\newcommand{\I}{\textup{i}}
\newcommand{\E}{\textup{e}}
\newcommand{\D}{\textup{d}}

\newcommand*{\und}{\ensuremath{\quad\text{and}\quad}}

\newcommand{\dd}{\text{d}}
\newcommand{\dod}[2]{\frac{\dd #1}{\dd #2}}

\newcommand{\pop}[2]{\frac{\partial #1}{\partial #2}}

\newcommand{\ddim}{\udelta\kern0.1em}

\newcommand{\beikonst}[2]{\left( #1 \right)_{\kern-0.2em #2}}

\newcommand{\intvar}[2][]{\D^{#1}\kern-0.2em #2\,}

\newcommand{\anderStelle}[2]{\left. #1 \right|_{#2}}

\newcommand*{\Abs}[1]{\left|#1\right|}

\newcommand{\VolProb}[2][]{\mathcal{V}^{#1} \kern-0.2em \left( #2 \right)}

\newcommand*{\bra}[1]{\mathopen{\langle}#1\mathclose{|}}
\newcommand*{\ket}[1]{\mathopen{|}#1\mathclose{\rangle}}

\newcommand{\psioben}{\psi^{\text{ex}}}
\newcommand{\psiunten}{\psi^{\text{gr}}}

\newcommand{\Entartung}[3][]{N^{\text{#2}}_{#3} #1}

\newcommand{\Entunten}{\Entartung{c}{1}}
\newcommand{\Entoben}{\Entartung{c}{0}}

\newcommand{\wahr}[3][]{W^{#2}_{#3} #1}

\newcommand{\wahroben}{\wahr{\text{ex}}{}}
\newcommand{\wahrunten}{\wahr{\text{gr}}{}}

\newcommand{\Energie}[3][]{E^{#2}_{#3} #1}
\newcommand{\Energy}{E}

\newcommand{\Energyn}[1]{\Energie{}{#1}}

\newcommand{\breite}{\delta\epsilon} 
\newcommand{\DeltaEoben}{\Delta\Energy^{\text{c}}_0} 
\newcommand{\DeltaEunten}{\Delta\Energy^{\text{c}}_1} 

\newcommand{\op}[1]{\hat{#1}}
\newcommand{\Hamiltonian}{\hat{H}}

\newcommand{\WWways}{\hat{V}}

\newcommand{\durchmat}{\lambda_0} 

\newcommand{\1}{\op{1}}
\newcommand{\Projektor}[1][]{\op{P}#1}
\newcommand{\Poben}{\Projektor^{\text{ex}}}
\newcommand{\Punten}{\Projektor^{\text{gr}}}

\newcommand{\DiracTrafo}[1]{\hat{U}_{#1}}

\newcommand{\timeopeins}{\hat{U}_1}

\newcommand{\timeopzwei}{\hat{U}_2}

\newcommand*{\sprod}[2]{\mathopen{\langle}#1 | #2\mathclose{\rangle}}

\newcommand{\tr}[2][]{\text{Tr}_{#1}\left\{#2\right\}}
\newcommand{\trtxt}[2][]{\text{Tr}_{#1}\{#2\}}

\newcommand*{\Haver}[1]{\mathopen{\llbracket} #1 \mathclose{\rrbracket}}

\begin{document} 

%
%
\begin{frontmatter}

\title{From Pure Schr\"odingerian to Statistical Dynamics } 

\author[osn]{Jochen Gemmer\corauthref{cor}} and
\corauth[cor]{Corresponding author.}
\ead{jgemmer@uos.de}
\author[stg]{Mathias Michel}

\address[osn]{Fachbereich Physik, Universit\"at Osnabr\"uck, 49069 Osnabr\"uck, Germany}
\address[stg]{Institut f\"ur Theoretische Physik I, Universit\"at Stuttgart, 70550 Stuttgart, Germany}

\date{\today}%
\begin{abstract}
Many processes in nature seem to be entirely controlled by transition 
rates and the corresponding statistical dynamics. Some of them are in
essence quantum, like the decay of excited states, the tunneling
through barriers or the decay of unstable nuclei. Thus, starting from
first principles, those systems should be analyzed on the basis of the
Schr\" odinger equation. In the present paper we consider a two level
system coupled to an environment which is basically described by an
two-band energy scheme. For appropriately tuned environment
parameters, the excitation probability of the two level system exhibits
statistical dynamics, while the full system follows the coherent,
unitary pure state evolution generated by the Schr\"odinger equation.
\end{abstract}

\end{frontmatter}

%
%

\section{Introduction}

Quite naturally the question for the decay behavior of quantum states
is not new and thus there is a considerable amount of theories aiming
in that direction. 

First of all Fermi's Golden Rule should be
mentioned, a formula of enormous practical importance. Nevertheless
its derivation is based on the assumption of a short perturbation and
thus in can hardly account for a full, continuous decay process, all
the way down to equilibrium.

The decay process associated with the spontaneous emission of an atom is
described by the Weisskopf Wigner Theory in a way that is not based
on a short perturbation. But this theory describes particularly an
atom coupled to the electromagnetic field in the vacuum state and is
thus hard to generalize to arbitrary environments in arbitrary states.

In the context of the system-environment scenario there are also the theories based on quantum master
equations. Most of them involve some projection operator technique as
well as the  Born approximation and typically systems must be
Markovian \cite{Kubo1991}. Many
derivations are based on an initial state which is a product state
with a thermal bath part. Some derivations even assume this structure
for the full relaxation process \cite{Scully1997,Blum1996}. This assumption is often backed
up by the argument that since for no interaction no system-environment
correlations could be generated, weak interactions (Markovicity) could
only generate negligible correlations \cite{Pechukas1994,Weiss1999}. Furthermore, since those master
equations result in maps on the considered system of the form
$\hat{\rho}(t+\tau)=\mathcal{M}(\tau)\hat{\rho}(t)$ that allow for an
iteration as
$\mathcal{M}(\tau_1+\tau_2)=\mathcal{M}(\tau_1)\mathcal{M}(\tau_2)$ \cite{Alicki1987},
the final state with regard to one map might be the initial state with
regard to another. This fact also seems to support the idea that at
least weak interactions could not produce considerable correlations.

In the following we  thus shortly comment on the question of
relaxation and correlations in general, and then introduce an
alternative theory of the relaxation process that takes arbitrary
pure, possibly correlated or even entangled full system states into account.

\section{Relaxation and Correlations}

Is it possible that a full bi-partite system (system-reservoir-model)
undergoes a unitary transformation, such that the purity of the
considered system decreases
(entropy increases, relaxation), without substantial system-reservoir correlations being generated?

This question shall be addressed within this section. To those ends we
specify the ``correlations'', $\hat{\rho}_c$ as an addend of the full
system density matrix, $\hat{\rho}$:
\begin{equation}
\hat{\rho}_c:=\hat{\rho}-\hat{\rho}_s \otimes \hat{\rho}_r \quad \left(\hat{\rho}_s:=\mbox{Tr}_r\left\{\hat{\rho}\right\}, \hat{\rho}_r:=\mbox{Tr}_s\left\{\hat{\rho}\right\}\right)
\end{equation}
Obviously $\hat{\rho}_s \otimes \hat{\rho}_r$ specifies the
uncorrelated product part of the density matrix.
To measure the ``size'' of the correlated and the uncorrelated parts
we use the absolute value, $P$, of an operator: 
\begin{equation}
P_x:=\sqrt{\mbox{Tr}\left\{\hat{\rho}^2_x\right\}}\quad x=s,r,\mbox{none}
\end{equation}
Evidently $P$ is also the purity of the corresponding (sub)system. To
decide whether or not correlations are negligible we finally want to consider
the ``correlations/product'' coefficient $\eta$: 
\begin{equation}
\eta:= \frac{P_c}{P_sP_r}
\end{equation}
If $\eta \ll 1$, correlations may safely be neglected.

Computing the size of the correlations yields:
\begin{equation}
\label{sico}
P_c^2=P^2-2\mbox{Tr}\left\{\hat{\rho} \hat{\rho}_s \hat{\rho}_r\right\}+P_s^2P_r^2
\end{equation}
Since the trace of a product of two hermitian matrices fulfils the
conditions on an inner product, one finds through application of
Schwartz' s inequality:
\begin{equation}
|\mbox{Tr}\left\{\hat{\rho} \hat{\rho}_s \hat{\rho}_r\right\}| \leq PP_sP_r
\end{equation}
Inserting this into (\ref{sico}) yields
\begin{equation}
\label{sico1}
P_c^2 \geq (P-P_sP_r)^2
\end{equation}
or for the coefficient $\eta$ 
\begin{equation}
\label{seta}
 \eta \geq \frac{P}{P_sP_r}-1
\end{equation}
$P$ is invariant under unitary transformation. Often the reservoir is
assumed to be exactly stationary which might not precisely hold true,
nevertheless $P_r(0)\approx P_r(t)$ should be a reasonable
approximation for large reservoirs. Thus the only quantity that may
substantially change upon relaxation on the right hand side of
(\ref{seta}) is $P_s$. And if  $P_s$ decreases, $\eta$ obviously
increases. For the case of a stationary bath and an initial product
state one finds 
\begin{equation}
\eta \geq \frac{P_s(0)}{P_s(t)}-1
\end{equation}
This lower bound on $\eta$ may easily take on rather high values,
{\it e.g.}, for an $N$-level system coupled to a bath in the high
temperature limit ($kT$ much larger than the level spacing) one gets for an initially pure state, $P_s(0)=1$,
\begin{equation}
\eta \geq \sqrt{N}-1
\end{equation}
This result is absolutely independent of the interaction strength. It
only connects a decrease of purity (increase of entropy) to an
increase of system-reservoir correlations, regardless of the
timescale on which this relaxation process happens. 
Thus we conclude that, quite contrary to the idea of system and bath
remaining uncorrelated correlations
are generically generated upon relaxation.

\section{The Hilbert Space Average Method}
\label{ham}

 The method of describing the relaxation process we are going to present here relies on a \stepwise\
 solution of the
 \sgleichung . Naturally, this approach, like all other approaches, relies on approximations.

The crucial approximation here is the replacement of some specific
quantities by their {\hsav}s. {\hsav}s are averages of quantities
defined as functions of the pure full system state, over sets of such
states which share a crucial common feature, like, {\it e.g.}, in our
case, the same excitation probability for the considered system. If the
distribution of excitation probabilities among the states of the
respective set is broad, the replacement of an actual value by its
{\hsav} will most likely be a bad approximation. If the distribution
is tightly centered around its mean value, the replacement still only represents a ``best unbiased
guess'' but with distributions getting narrower, this guess obviously
gets better and better. In our case there is evidence that the width
of the distribution vanishes with an increasing number of environment
states.

Based on this idea we derive a rate equation, that describes the decay
which is controlled only by the Schr\"odinger equation, up to some fluctuations.

%
%
\section{The System-Environment Model}
\label{chap:ways:sec:envapp}
\index{system}
\index{environment}

\begin{figure}
\centering
\psfrag{eins}{$\ket{0}$}
\psfrag{null}{$\ket{1}$}
\psfrag{dE}{$\Delta\Energy$}
\psfrag{otimes}{$\bigotimes$}
\psfrag{I}{\vspace{3mm}$\hat{V}$}
\psfrag{N1}{$\Entunten$}
\psfrag{N0}{$\Entoben$}
\psfrag{gas}{\hspace{3mm}\qmarks{gas} g}
\psfrag{container}{\hspace{4mm}\qmarks{container} c}
\psfrag{deps}{$\breite$}
\includegraphics[width=6.5cm, height=3cm]{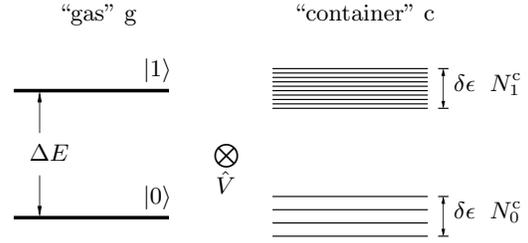}
\caption{Discrete two-level system coupled to a quasi-continuous container system. This \setup\ exhibits, for a sufficiently high state density in the container system, and an adequately tuned coupling, an exponential decay of an excitation in the gas system.}
\label{fig:ways:sec:envapp:1}
\end{figure}

The energy scheme of the situation we are going to analyze is depicted in \reffig{fig:ways:sec:envapp:1}.
A two level system, g (``gas''), is in contact with a \qmarks{many
  level} environment or \qmarks{container}, c (We use this
nomenclature for purely historical reasons \cite{Gemmer2001}. 
Only the relevant parts of the spectrum of the environment enter the model. 
These are, in this case, two \qmarks{bands} of width $\breite$, containing $\Entunten$ $(\Entoben)$ equidistant eigenstates in the upper (lower) band.
Therefore the level spacing within the upper (lower) energy \qmarks{band} is 
\begin{equation}
  \label{eq:322}
  \DeltaEunten := \frac{\breite}{\Entunten} \und
  \DeltaEoben  := \frac{\breite}{\Entoben}\;.
\end{equation}
In the following, quantities of the \qmarks{upper band} of the environment get the subscript 1, where as quantities of the \qmarks{lower band} get the subscript 0.
If we, {\it e.g.}, consider an evolution from an initial state, with the system in the excited state $\ket{1}$ and the environment in the \qmarks{lower band}, 0, due to overall \energyconservation\ the only other set of states that the full system can evolve into, is the set with the considered system in the ground state $\ket{0}$ and the environment in its \qmarks{upper band}, 1. 
The \hamiltonian\ within the relevant subspace of the entire \hilbertspace\ may thus be organized as follows,
\begin{equation}
  \label{eq:269}
  \Hamiltonian = \left( %
  \begin{array}{ccc|ccc} %
  \ddots &                & 0      &          &              &        \\ %
         & i\DeltaEunten  &        &          & \WWways      &        \\ %
  0      &                &\ddots  &          &              &        \\ %
  \hline 
         &                &        & \ddots   &              & 0      \\ %
         & \WWways        &        &          & j\DeltaEoben &        \\ %
         &                &        & 0        & \ddots       &        \\ %
  \end{array} \right) %
  \begin{array}{cc} %
  \left. %
  \begin{array}{c}
  \\ \\ \\
  \end{array}
    \right\} & \ket{\psiunten}  \\
  \left.
  \begin{array}{c}
  \\ \\ \\
  \end{array}
    \right\} & \ket{\psioben} \\
  \end{array}
\end{equation}
where $i$ ($j$) count the levels in the upper (lower) \qmarks{band} of the environment.
The \hamiltonian\index{Hamiltonian}\index{operator!Hamilton}\index{Hamilton operator} is displayed in the eigenbasis of the uncoupled system, for simplicity we assume for the moment that the coupling $\WWways$ only adds terms to the \hamiltonian\ in the \offdiagonal\ blocks.
This corresponds to an coupling which may give raise to energy
exchange between the subsystems. 

We now introduce two projectors\index{projector}, which project out the upper (lower) part of the state of the system
\begin{equation}
  \label{eq:324}
  \Poben :=\ket{1}\bra{1}\otimes\1^{(\text{c})}\;, \quad%
  \Punten:=\ket{0}\bra{0}\otimes\1^{(\text{c})}\;,
\end{equation}
where $\1^{(\text{c})}$ is the \einsop\ in the environmental system.
In the following we call that part of the wave vector that corresponds to the considered system in the excited state\index{excited state}\index{state!excited} $\ket{\psioben}$ and the part that corresponds to the system in the ground state\index{ground state}\index{state!ground} $\ket{\psiunten}$, \ie,
\begin{align}
  \label{eq:270}
 &\ket{\psioben} :=\Poben \ket{\psi}, \;\ket{\psiunten}:=\Punten\ket{\psi}\\
 &\Rightarrow \ket{\psi} = \ket{\psioben}+\ket{\psiunten}\nonumber.
\end{align}
Note that neither $\ket{\psioben}$ nor $\ket{\psiunten}$ are normalized individually.

To analyze this model we first transform to the Dirac or interaction
picture \cite{Schiff1968}
\begin{align}
  \label{eq:271}
  &\op{U}(t,0) := \DiracTrafo{0} 
              := \E^{-\frac{\I}{\hbar}\Hamiltonian_0 t}\;, \quad %
  \ket{\psi_{\text{I}}}:=\DiracTrafo{0}^{\dagger}\ket{\psi}\;, \quad \\
  &\WWways_{\text{I}} := \DiracTrafo{0}^{\dagger}\,\WWways\,\DiracTrafo{0},\n
\end{align}
where $\Hamiltonian_0$ is the \hamiltonian\ of the uncoupled system.
The \sgleichung\ in this representation reads
\begin{equation}
  \label{eq:272}
  \I \hbar\, \pop{}{t} \ket{\psi_{\text{I}}} = %
  \WWways_{\text{I}} \ket{\psi_{\text{I}}}\;,
\end{equation}
where both states and operators are now \timedependent\index{state!time-dependent}\index{operator!time-dependent}, \ie\ also $\WWways_{\text{I}}$ is a time dependent operator, but preserves the \offdiagonal\ block form as before.

The crucial quantities in the context of a decay\index{decay} to equilibrium\index{equilibrium} are the probabilities\index{probability} to find the system in its exited (ground) state, $\wahroben\, (\wahrunten)$. 
Due to the diagonality of $\Hamiltonian_0$ those quantities have the same representation in the interaction as well as in the \sbild\index{Schrodinger@Schr\"odinger!picture},
\begin{align}
  \label{eq:273}
  &\wahroben  = \sprod{\psioben_{\text{I}}}{\psioben_{\text{I}}}
             = \sprod{\psioben}{\psioben}\\
  &\wahrunten = \sprod{\psiunten_{\text{I}}}{\psiunten_{\text{I}}} 
             = \sprod{\psiunten}{\psiunten} \;.\n
\end{align}
For simplicity we omit in the following the \ibild\index{interaction!picture} subscript \qmarks{I}, but all the following considerations refer to this picture.

%
%
\section{Time Evolution}
\label{chap:ways:sec:time}
\index{evolution!time}

To approximate the evolution of the system for a short \timestep, we can truncate the corresponding Dyson series
\begin{equation}
  \label{eq:274}
  \ket{\psi(\tau)} \approx \left(\1 - \frac{\I}{\hbar}\,
  \timeopeins(\tau) - \frac{1}{\hbar^2}\,\timeopzwei(\tau)\right)\ket{\psi(0)}
  \;.
\end{equation}
This is a truncation of second order, in which the $\hat{U}$'s are the time ordered integrals that occur in the \name{Dyson} series \cite{Schiff1968}
\begin{align}
  \label{eq:275}
  &\timeopeins(\tau) = \int_0^{\tau}\D\tau'\WWways(\tau')\;,\\
  &\timeopzwei(\tau) = \int_0^{\tau}\D\tau'\WWways(\tau')\int_0^{\tau'}\D\tau''
  \WWways(\tau'')\;.\n
\end{align}
According to the hermiticity of $\WWways(\tau)$, $\timeopeins(\tau)$ should be hermitian too, which is not the case for $\timeopzwei(\tau)$.
$\timeopeins(\tau)$ has the same \offdiagonal\ form as $\WWways(\tau)$ whereas $\timeopzwei(\tau)$ has here a block diagonal form according to the interaction matrix. 
(To further simplify notation we do not write the $\tau$ dependence of the $\hat{U}$'s explicitly.
Furthermore, we omit the time dependence of the wave function, if it refers to the initial state, \ie, $\ket{\psi(0)} := \ket{\psi}$). 

As mentioned above we are interested in the time evolution of the probability to find the system in its excited state $\wahroben(\tau)$, or ground state $\wahrunten(\tau)$, respectively.
Initially we consider $\wahroben(\tau)$.
Neglecting all terms of higher than second order (products of
$\timeopeins$ and $\timeopzwei$ as well as terms proportional to
$\timeopzwei^2$) and taking the special \offdiagonal\ block form of
the interaction into account, we get from (\ref{eq:274}) and (\ref{eq:270})
\begin{align}
 \label{eq:276}  
 \wahroben(\tau) %
    &=  \sprod{\psioben}{\psioben} %
     + \frac{\I}{\hbar}\,\bra{\psiunten}\timeopeins\ket{\psioben} \\%
    &- \frac{\I}{\hbar}\,\bra{\psioben} \timeopeins\ket{\psiunten}\notag\\
    &+ \frac{1}{\hbar^2}\,\bra{\psiunten}\timeopeins^2\ket{\psiunten} %
     - \frac{1}{\hbar^2}\,\bra{\psioben}(\timeopzwei %
                 +\timeopzwei^{\dagger})\ket{\psioben} \notag\\
  \label{eq:277}
  \wahrunten(\tau) %
    &=  \sprod{\psiunten}{\psiunten} %
     + \frac{\I}{\hbar}\,\bra{\psioben} \timeopeins\ket{\psiunten} \\%
    &- \frac{\I}{\hbar}\,\bra{\psiunten}\timeopeins\ket{\psioben} \notag\\
    &+ \frac{1}{\hbar^2}\,\bra{\psioben}\timeopeins^2\ket{\psioben} %
     - \frac{1}{\hbar^2}\,\bra{\psiunten}(\timeopzwei %
                 +\timeopzwei^{\dagger})\ket{\psiunten}\notag
\end{align}%
The strict overall probability conservation requires
\begin{align}
  \label{eq:278}
  &\wahroben(\tau)+\wahrunten(\tau)\notag\\
  &= \sprod{\psioben(\tau)}{\psioben(\tau)} + %
     \sprod{\psiunten(\tau)}{\psiunten(\tau)} \notag\\%
  &= \sprod{\psioben}{\psioben} + \sprod{\psiunten}{\psiunten} = 1\;.
\end{align}
Since the normalization is already fulfilled in the zero order, all higher orders must vanish.
Obviously the first order vanishes automatically. 
Thus, exploiting (\ref{eq:278}), for the second order of the sum of
(\ref{eq:276}) and (\ref{eq:277}) we find
\begin{align}
  \label{eq:279}
  \bra{\psioben}(\timeopzwei^{\phantom{\dagger}}
                 +\timeopzwei^{\dagger})\ket{\psioben} &= %
  \bra{\psioben}\timeopeins^2\ket{\psioben} %
  \,,\\ %
  \bra{\psiunten}(\timeopzwei^{\phantom{\dagger}}
                  +\timeopzwei^{\dagger})\ket{\psiunten} &= %
  \bra{\psiunten}\timeopeins^2\ket{\psiunten}\;.\n
\end{align}
Inserting this into (\ref{eq:276}) and (\ref{eq:277}) yields
\begin{align}
  \label{eq:280}
  \wahroben(\tau) &= \sprod{\psioben}{\psioben} %
     +\frac{\I}{\hbar}\bra{\psiunten}\timeopeins\ket{\psioben} \notag\\ 
   & -\frac{\I}{\hbar}\bra{\psioben}\timeopeins\ket{\psiunten} \notag\\
   & + \frac{1}{\hbar^2}\bra{\psiunten}\timeopeins^2\ket{\psiunten} %
     -\frac{1}{\hbar^2}\bra{\psioben}\timeopeins^2\ket{\psioben} \\
  \wahrunten(\tau) &= \sprod{\psiunten}{\psiunten} %
     +\frac{\I}{\hbar}\bra{\psioben}\timeopeins\ket{\psiunten} \notag\\
   & -\frac{\I}{\hbar}\bra{\psiunten}\timeopeins\ket{\psioben} \notag\\
   & + \frac{1}{\hbar^2}\bra{\psioben}\timeopeins^2\ket{\psioben} %
     -\frac{1}{\hbar^2}\bra{\psiunten}\timeopeins^2\ket{\psiunten}\;.
\end{align}

For an exact evaluation of the right hand side one would need to know
the $\ket{\psioben},\ket{\psiunten}$ in detail. But rather than doing
so, we replace now the actual quantities by their corresponding Hilbert space averages, according to the
approximation scheme explained in sect.\ref{ham}. The appropriate set of
states over which the average has to be taken here is the set of all states
featuring the same excitation probability
$\sprod{\psioben}{\psioben}$. Some justification for this replacement
has been given in sect.\ref{ham}, some comes from the numerical
results, sect. \ref{num}, and for full detailed justification, see
\cite{Gemmer2003,Gemmer2004}.
Since not only the justification of the replacement but also the
actual computation of the respective are beyond the scope of
this text we again refer the interested reader to \cite{Gemmer2003,Gemmer2004}. Here, we
only want to give and discuss the results:
\begin{align}
  \label{eq:472}
  &\Haver{\bra{\psiunten}\timeopeins\ket{\psioben}}=\Haver{\bra{\psioben}\timeopeins\ket{\psiunten}} = 0 \;,\\
&\Haver{\bra{\psioben}\timeopeins^2\ket{\psioben}}
                  = \frac{\sprod{\psioben}{\psioben}}{\Entoben}\, %
                    \trtxt[\text{ex}]{\timeopeins^2} \;, \n\\
 &\Haver{\bra{\psiunten}\timeopeins^2\ket{\psiunten}}
                   = \frac{\sprod{\psiunten}{\psiunten}}{\Entunten}\, %
                     \trtxt[\text{gr}]{\timeopeins^2} \;,\n 
\end{align}
where $\Haver{\cdots}$ denotes the Hilbert space average and $\trtxt[\text{ex}(\text{gr})]{\dots}$ the trace\index{trace} over the upper (lower) subspace of the operator. 

Plugging those results into (\ref{eq:280}), we get
\begin{align}
  \label{eq:285}
  &\wahroben(\tau)= \\
  &\wahroben(0)+ \frac{\wahrunten(0)}{\hbar^2 \Entunten}\,
      \trtxt[\text{gr}]{\timeopeins^2} 
    - \frac{\wahroben(0)}{\hbar^2 \Entoben}\,
      \trtxt[\text{ex}]{\timeopeins^2}\;, \n\\
  \label{eq:286}
  &\wahrunten(\tau) = \n\\
  &\wahrunten(0)+\frac{\wahroben(0)}{\hbar^2 \Entoben}\,
     \trtxt[\text{ex}]{\timeopeins^2}
    -\frac{\wahrunten(0)}{\hbar^2 \Entunten}\,
     \trtxt[\text{gr}]{\timeopeins^2}\;.\n
\end{align}
Now we have to analyze those traces in more detail. 
We will do this explicitly for the upper subspace but by simply exchanging the indices, the result will be valid for the lower subspace as well
\begin{equation}
  \label{eq:287}
  \trtxt[\text{ex}]{\timeopeins^2} 
  = \sum_{j=1}^{\Entoben}\bra{j}\timeopeins^2\ket{j} 
  = \sum_{j=1}^{\Entoben}\Abs{\timeopeins\ket{j}}^2\;.
\end{equation}
Here $j$ runs over the eigenstates of $\hat{H}_0$ in the upper subspace (note this corresponds to the lower \qmarks{band} of the environment). 
The object that is summed over here is evaluated in the literature in the context of \FgoldenRule,
\begin{equation}
  \label{eq:288}
  \Abs{\timeopeins\ket{j}}^2 = %
  \sum_{i=1}^{\Entunten} \Abs{\bra{i}\WWways\ket{j}}^2 %
      \frac{4 \sin^2(\frac{1}{2}\,\omega_{i,j}\,\tau)}{\omega^2_{i,j}}
\end{equation}
with
\begin{equation}
  \label{eq:289}
  \omega_{i,j} %
     = \frac{1}{\hbar}\left(\Energyn{j}-\Energyn{i}\right)   
     = \frac{1}{\hbar}\left( j \DeltaEoben - i \DeltaEunten\right)\;,
\end{equation}
see \reffig{fig:ways:sec:step:2a}.

%
%
\section{The Linear Regime}
\label{chap:ways:sec:step}
\index{time!short step}

Our arguments, including the conditions we have to impose on the model, follow now closely the ones brought forth in the context of \FgoldenRule.

\begin{figure}
\centering
\psfrag{f}{$f(\omega)$}
\psfrag{t}{$\frac{\tau^2}{4}$}
\psfrag{A}{$A=\pi\tau/2$}
\psfrag{dw}{$\delta\omega$}
\psfrag{w}{$\omega$}
\includegraphics[width=6cm]{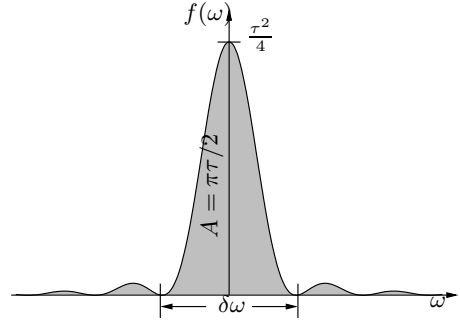}
\caption{The function $f(\omega)$ defined in (\ref{eq:290}).}
\label{fig:ways:sec:step:1}
\end{figure}

\begin{figure}
\centering
{\psfrag{de}{$\breite$}
 \psfrag{W}{$\omega_{i,1}$}
 \psfrag{Ne}{$\Entoben = 4$}
 \psfrag{Ng}{$\Entunten = 7$}
 \psfrag{m1}{$\scriptstyle j=1$}
 \psfrag{m2}{$\scriptstyle \phantom{j=}2$}
 \psfrag{n1}{$\scriptstyle i=1$}
 \psfrag{n2}{$\scriptstyle  \phantom{i=}2$}
 \subfigure[]{\includegraphics[width=4.5cm, height=3cm]{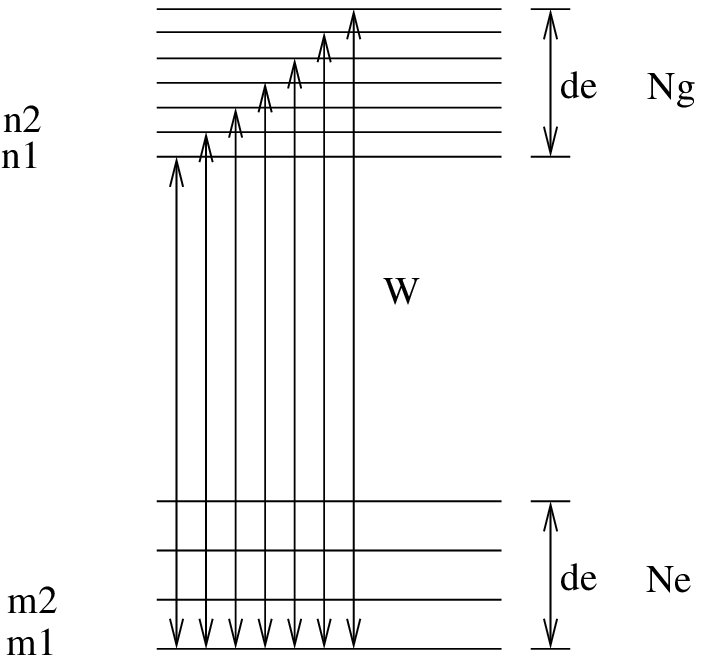}
              \label{fig:ways:sec:step:2a}}
}\\
{
 \psfrag{d}{$\delta\omega \gg \delta\omega(\tau_1)$}
 \psfrag{f}{$f(\omega)$}
 \psfrag{w}{$\omega$} 
 \psfrag{t}{$\tau \ll \tau_1$}
 \subfigure[]{\includegraphics[width=4.5cm]{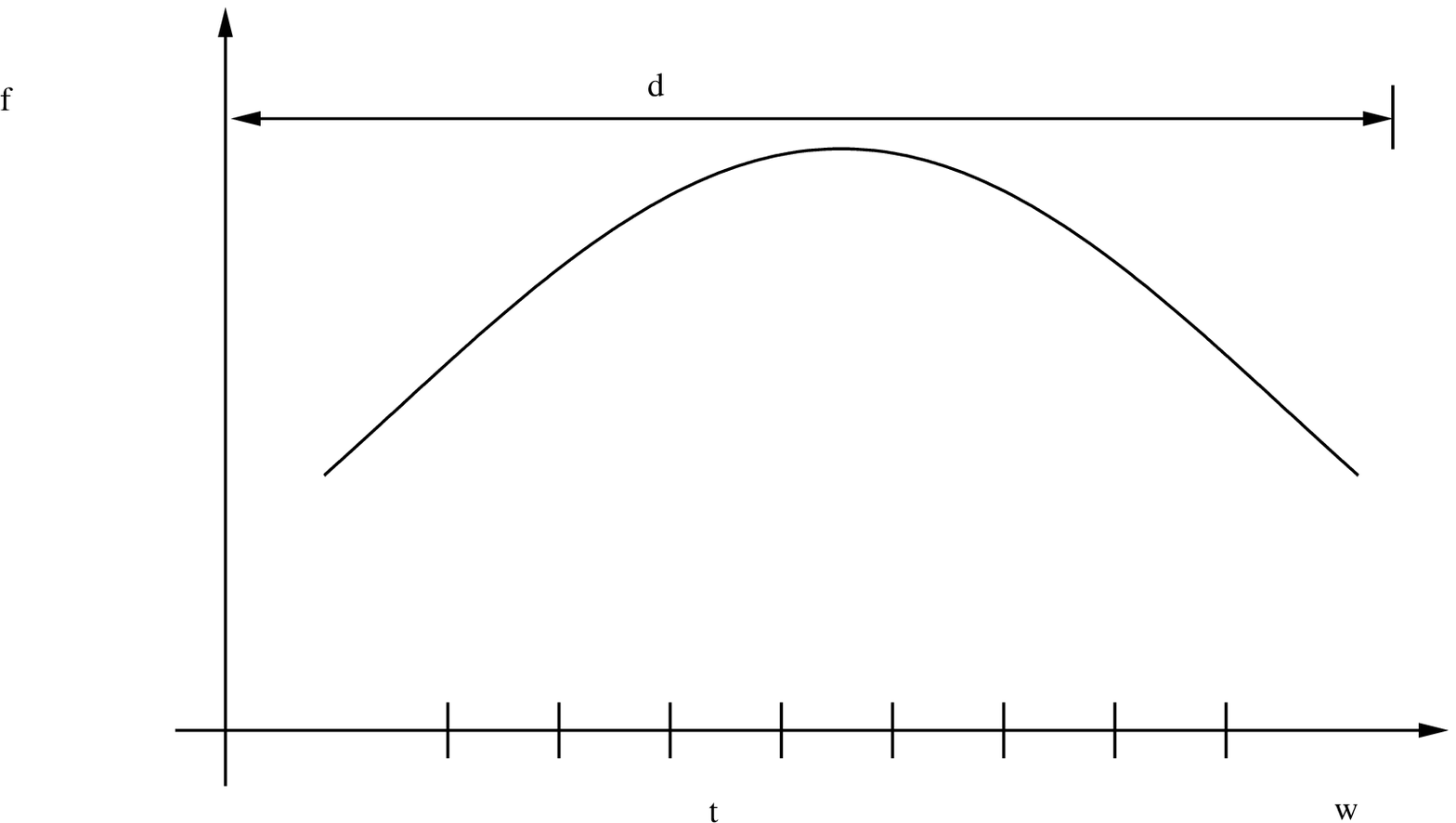}
              \label{fig:ways:sec:step:2b}}
}\\
{\psfrag{dw}{$\delta\omega(\tau_1)$}
 \psfrag{f}{$f(\omega)$}
 \psfrag{w}{$\omega$}
 \psfrag{delta}{$\scriptstyle \Delta\omega$}
 \psfrag{n}{\hspace{-2mm}$\scriptstyle i$}
 \psfrag{n=1}{\tiny 1}
 \psfrag{2}{\hspace{-1mm}\tiny 2}
 \psfrag{3}{\hspace{-1mm}\tiny 3}
 \psfrag{4}{\hspace{-1mm}\tiny 4}
 \psfrag{5}{}
 \subfigure[]{\includegraphics[width=4.5cm]{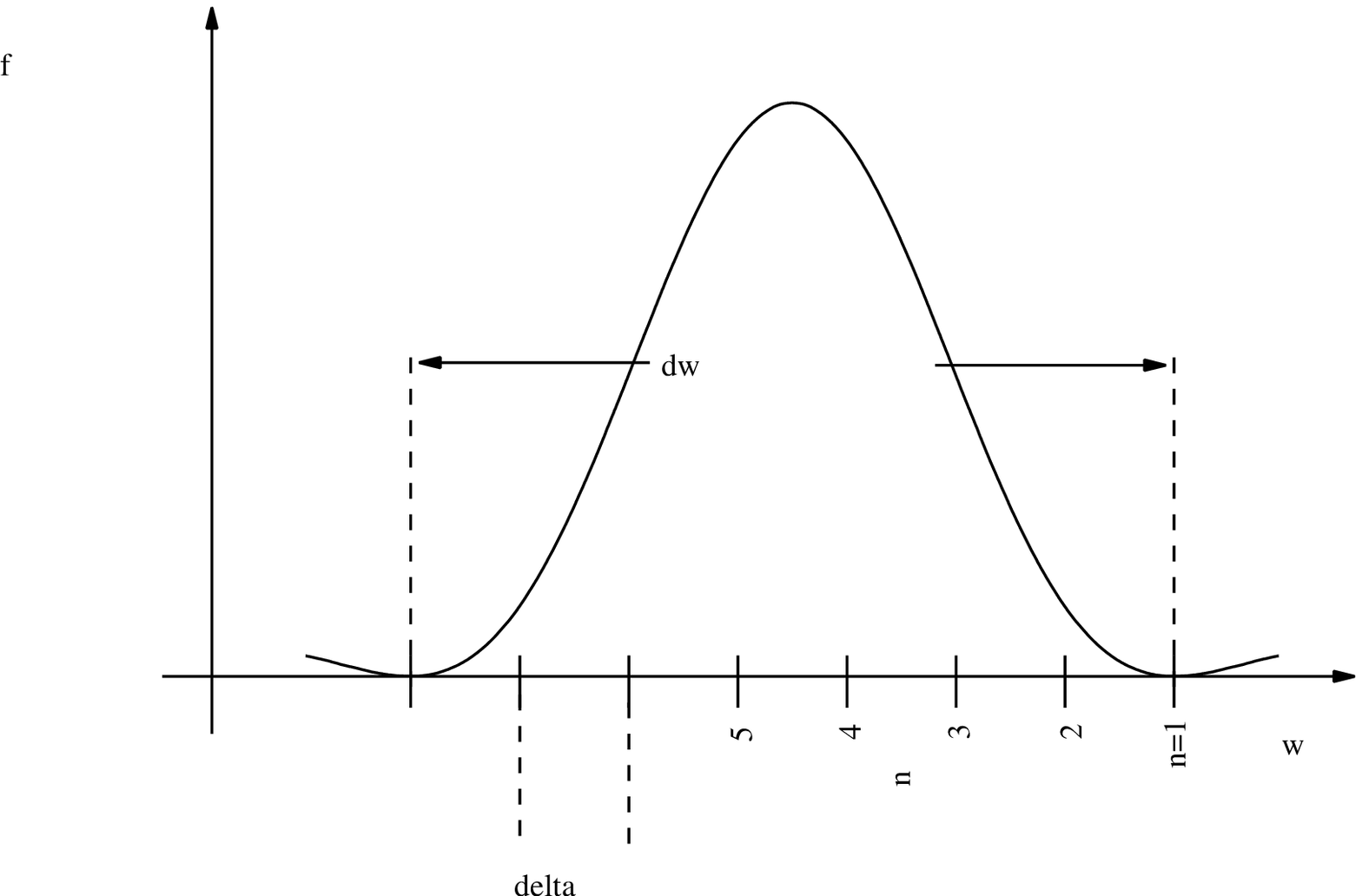}
              \label{fig:ways:sec:step:2c}}
}\\
{\psfrag{t}{$\tau=\tau_2$}
 \psfrag{f}{$f(\omega)$}
 \psfrag{a}{  }
 \psfrag{w}{$\omega$}
 \subfigure[]{\includegraphics[width=4.5cm]{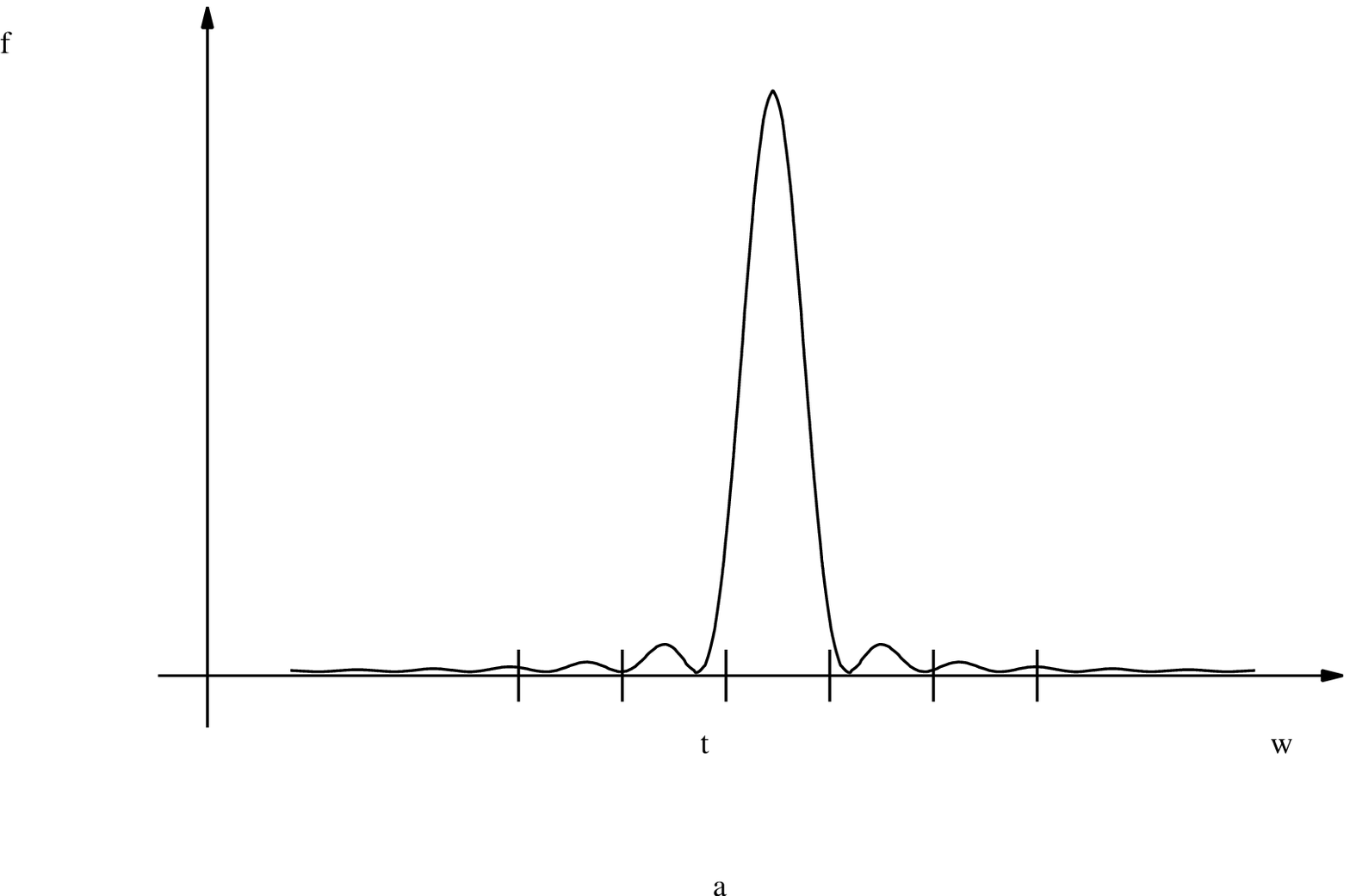}
              \label{fig:ways:sec:step:2d}}
}
\caption{Summation of transitions in (\ref{eq:288}): (\textbf{a}) Matrix elements to be summed up. (\textbf{b}) $\tau \ll \tau_1$: almost all terms are around the maximum of the peak (square regime). (\textbf{c}) $\tau \approx \tau_1$: the terms are distributed over the whole peak (linear regime). (\textbf{d}) $\tau \approx \tau_2$: only a few terms are within the peak (break-down of the approximation).}
\label{fig:ways:sec:step:2}
\end{figure}

The summation in (\ref{eq:288}) consists of two different terms: the
transition elements of the interaction matrix and a weight $f(\omega)$.
The spacing of different $\omega_{i,j}$ is given by
\begin{equation}
  \label{eq:329}
  \Delta \omega = \omega_{i,j}-\omega_{i+1,j} %
                = \frac{\DeltaEunten}{\hbar} %
                = \frac{\breite}{\Entunten\hbar}\;,
\end{equation}
where we have used (\ref{eq:322}).
The function
\begin{equation}
  \label{eq:290}
  f(\omega) = \frac{\sin^2(\frac{1}{2} \omega\tau)}{\omega^2}
\end{equation}
is basically a peak at $\omega=0$, with the width $\delta\omega=4\pi/\tau$ and a height of $f(0)=\tau^2/4$. 
The area under the function $f(\omega)$ is $A=\pi\tau/2$ (see \reffig{fig:ways:sec:step:1}).
This means the peak gets higher and narrower as $\tau$ increases (see \reffig{fig:ways:sec:step:2}).

The height of the peak grows with the square of the time $\tau$, the area under $f$ only linearly with $\tau$.
One could thus expect two different behaviors: the square- and the linear regime\index{regime!square}\index{regime!linear}. 
At the very beginning, the peak is very broad and therefore much broader than the \qmarks{band} width $\breite$ divided by $\hbar$. 
In this case we expect that the sum grows with the square of $\tau$, because all terms are near the maximum of the peak (see \reffig{fig:ways:sec:step:2b}).
We choose some $\tau_1$ such that the width $\delta\omega(\tau_1)$ of $f(\omega)$ has approximately the same value as the \qmarks{band} width $\breite$ divided by $\hbar$
\begin{equation}
  \label{eq:291}
  \delta\omega(\tau_1) = \frac{4\pi}{\tau_1} %
     \approx \frac{\breite}{\hbar} \quad \Rightarrow \quad %
  \tau_1 = \frac{4\pi\hbar}{\breite}\;.
\end{equation}
The terms are distributed over the whole width of the peak and we expect that the sum grows proportional to the area under the peak, thus linearly in $\tau$ (see \reffig{fig:ways:sec:step:2c}) .
In this case and if, furthermore, the function $f$ does not change much over many summation steps $\Delta\omega$, \ie, if
\begin{equation}
  \label{eq:292}
  \Delta \omega = \frac{\breite}{\Entunten\hbar} \ll \delta\omega(\tau_1) %
                = \frac{\breite}{\hbar} %
  \quad \Rightarrow \quad  \Entunten \gg 1
\end{equation}
the summation averages out the different elements of the $\WWways$-matrix in (\ref{eq:288}). 
Therefore the sum may be approximated by the average of the interaction matrix element $\durchmat^2$ times the integral over $f(\omega)$ according to $\omega$. 
The average of the interaction matrix element is
\begin{equation}
  \label{eq:293}
  \durchmat^2 = \frac{1}{\Entunten\Entoben} %
                \sum_{i=1}^{\Entunten}\sum_{j=1}^{\Entoben} %
  \Abs{\bra{i}\WWways\ket{j}}^2 = \frac{1}{2\Entunten\Entoben} \trtxt{\WWways^2}\;.
\end{equation}
For (\ref{eq:288}) we then get
\begin{equation}
  \label{eq:294}
  \Abs{\timeopeins \ket{j}}^2 \approx %
  \durchmat^2 \int\frac{\D\omega}{\Delta\omega}\, 4 f(\omega) %
  = \frac{\durchmat^2 4A}{\Delta \omega} %
  = \frac{2\pi\durchmat^2 \hbar \Entunten \tau}{\breite}\;,
\end{equation}
where we have used that the area under $f(\omega)$ is $A=\pi\tau/2$, as mentioned before.

The approximation done so far breaks down later at some time $\tau_2$, when the peak gets too narrow (see \reffig{fig:ways:sec:step:2d}), \ie\ the width is smaller than the summation displacement $\Delta\omega$
\begin{equation}
  \label{eq:295}
  \delta\omega(\tau_2) = \frac{4\pi}{\tau_2} %
                       = \Delta\omega = \frac{\breite}{\Entunten\hbar}
  \quad \Rightarrow \quad \tau_2  = \frac{4 \pi\hbar\Entunten}{\breite}\;.
\end{equation}
Thus (\ref{eq:294}) is a valid approximation only for $\tau_1 < \tau <
\tau_2$, which is the linear regime.

Hence, plugging  (\ref{eq:294}) into (\ref{eq:287}) yields
\begin{equation}
  \label{eq:296}
  \tr[\text{ex}]{\timeopeins^2} 
  = \sum_{j=1}^{\Entoben}\Abs{\timeopeins\ket{j}}^2 %
  \approx \frac{2\pi\durchmat^2 \hbar\Entunten\Entoben\tau}{\breite}\;.
\end{equation}
Since this expression is symmetric under exchange of the upper and lower subspace, the corresponding expression for the lower subspace reads
\begin{equation}
  \label{eq:297}
  \tr[\text{gr}]{\timeopeins^2} 
  = \sum_{i=1}^{\Entunten}\Abs{\timeopeins\ket{i}}^2 %
  \approx \frac{2\pi\durchmat^2 \hbar\Entunten\Entoben\tau}{\breite}\;.
\end{equation}

Inserting (\ref{eq:296}) and (\ref{eq:297}) into (\ref{eq:285}) yields
\begin{align}
  \label{eq:298}
  \wahroben(\tau) &= \wahroben(0) + C\tau\Entoben\wahrunten(0)%
                                  - C\tau\Entunten\wahroben(0)\;,\\
  \wahrunten(\tau)&= \wahrunten(0) + C\tau\Entunten\wahroben(0) 
                                   - C\tau\Entoben\wahrunten(0)\;,\n
\end{align}
where we have abbreviated
\begin{equation}
  \label{eq:300}
  \frac{2\pi\durchmat^2}{\breite\,\hbar} := C\;.
\end{equation}
Equations (\ref{eq:298}) describe, within the discussed limits, a short \timestep\ starting from any initial state, not necessarily an eigenstate of $\Hamiltonian$. 
Since they directly connect the probabilities $\wahroben(0)$, $\wahrunten(0)$ of the initial state with those of the state reached after time $\tau$, we can now iterate these equations under some specific conditions. 

%
%
\section{Conditions on Model Parameters}
\label{chap:ways:sec:rateeq}
\index{rate!equation}
\index{equation!rate}

Before iterating the above equations (\ref{eq:298}), one should again check the {\precondition}s for the short \timestep\ equation derived so far. 
We have only considered terms up to second order, and we can only iterate after a \timestep\ of length $\tau_1$. 
Thus we have to make sure that the considered second order terms are still small compared to $1$ after $\tau_1$, to justify the dropping of higher order terms.
Therefore we must check that \eg\
\begin{equation}
  \label{eq:330}
  \anderStelle{C \tau \Entoben}{\tau=\tau_1} %
    = 8 \pi^2 \frac{\durchmat^2}{(\DeltaEoben)^2}\,\frac{1}{\Entoben}\ll 1\;,
\end{equation}
where we have used (\ref{eq:322}). 
In complete analogy we get for the other term of second order 
\begin{equation}
  \label{eq:301}
  \anderStelle{C \tau \Entunten}{\tau=\tau_1}
  = 8 \pi^2 \frac{\durchmat^2}{(\DeltaEunten)^2}\,\frac{1}{\Entunten}\ll 1\;.
\end{equation}
If these two conditions are fulfilled the \qmarks{linear regime}\index{regime!linear} is reached while the truncation to second order is still a valid description, and we can iterate (\ref{eq:298}) after some time $\tau>\tau_1$.
Obviously the linear regime is reached the faster the more levels the environment contains. 

However, if we want to use the above scheme (\ref{eq:298}) we should make sure that we iterate before the linear regime is left again, \ie, before $\tau_2$.
Therefore we must consider the second order terms at $\tau_2$ (\ref{eq:295}) compared to one. 
Note that $\tau_2$ differs for the two terms of second order, in (\ref{eq:295}) we only argued for one of the two energy \qmarks{bands} in the environment.
Thus, the case for which iterating (\ref{eq:298}) is the best description we can possibly get is
\begin{align}
  \label{eq:302}
  \anderStelle{C \tau \Entoben}{\tau=\tau_2} %
  &= 8 \pi^2 \frac{\durchmat^2}{(\DeltaEoben)^2} \geq 1\;, \\
  \label{eq:320}
  \anderStelle{C \tau \Entunten}{\tau=\tau_2} %
  &= 8 \pi^2 \frac{\durchmat^2}{(\DeltaEunten)^2} \geq 1\;.
\end{align}

\section{The Rate Equation}

If the above conditions are fulfilled, iterating (\ref{eq:298}) yields
\begin{align}
  \label{eq:303}
  &\frac{\wahroben((n+1)\tau)-\wahroben(n\tau)}{\tau}=\\
  &C \Entoben\wahrunten(n\tau)-C\Entunten\wahroben(n\tau)\;,\n\\
\n\\
  &\frac{\wahrunten((n+1)\tau)-\wahrunten(n\tau)}{\tau}=\n\\
  &C \Entunten\wahroben(n\tau)-C\Entoben\wahrunten(n\tau)\;.\n
\end{align}
Or, in the limit of $\tau$ being extremely small
\begin{align}
  \label{eq:305}
  \dod{\wahroben}{t} &= C \Entoben\wahrunten - C \Entunten\wahroben\;,\\
  \dod{\wahrunten}{t} &= C \Entunten\wahroben - C \Entoben\wahrunten\;.\n
\end{align}
This evolution equation for the probabilities obviously conserves the overall probability. 
We have obtained a rate equation for the probabilities to find the system in the upper respectively lower level.

The solutions of the equations (\ref{eq:305}) describe simple exponential decays\index{decay!exponential|textbf}, with exactly the same decay rates one would have gotten from \FgoldenRule.
A solution for the considered system being initially entirely in the exited state reads
\begin{align}
  \label{eq:307}
  \wahroben(t) &= \frac{\Entoben}{\Entunten+\Entoben} + 
                  \frac{\Entunten}{\Entoben+\Entunten}\, 
                  \E^{-C(\Entoben+\Entunten)t}\;, \\ 
  \wahrunten(t) &= \frac{\Entunten}{\Entoben+\Entunten}\, 
                   \left(1-\E^{-C(\Entoben+\Entunten)t}\right)\;.\n
\end{align}
The equilibrium values reached after very long times are
\begin{equation}
  \label{eq:308}
  \wahroben(\infty) = \frac{\Entoben}{\Entunten+\Entoben}\,, \quad %
  \wahrunten(\infty) = \frac{\Entunten}{\Entoben+\Entunten}\,,
\end{equation}
which are exactly the same as the ones derived in \cite{Gemmer2003} for the
equilibrium state of a system with an energy exchange coupling to a
possibly non-Markovian environment.

%
%
\section{Numerical Results for the Relaxation Period}
\label{num}

To check the validity of the theory developed in the previous Sections, a model of the type depicted in \reffig{fig:ways:sec:envapp:1}, with a \hamiltonian\ as described in (\ref{eq:269}) has been analyzed numerically by directly solving the \sgleichung. 
The interaction matrix $\WWways$ has been filled with random \Gaussian\index{matrix!random} distributed entries such that
\begin{equation}
  \label{eq:317}
  \frac{\durchmat^2}{(\DeltaEunten)^2} \approx 1\,,
\end{equation}
to ensure that (\ref{eq:330})-(\ref{eq:320}) are fulfilled. 
Different container sizes have been analyzed, corresponding to $\Entunten=$ 50, 100, 200, 400, 800 and $\Entoben=\frac{1}{2}\Entunten$. 
For all sizes the level spacings $\DeltaEunten$, $\DeltaEoben$ have been kept fixed such that for increasing container size the band widths increase.
 
With those parameter settings the theoretical prediction for
$\wahrunten(t)$ from (\ref{eq:307}) is the same for all container
sizes. The initial state is always chosen to be a pure product state
($S^g(0)=0$), with the gas-system in the ground state and the
container-system in a random superposition of states from the upper band. 

\begin{figure}
\centering
\psfrag{rho^g_00}{$\wahrunten$}
\psfrag{N_u=50}{\hspace{-5mm}$\scriptstyle \Entunten=50$}
\psfrag{N_u=100}{\hspace{-5mm}$\scriptstyle \Entunten=100$}
\psfrag{N_u=200}{\hspace{-5mm}$\scriptstyle \Entunten=200$}
\psfrag{N_u=400}{\hspace{-5mm}$\scriptstyle \Entunten=400$}
\psfrag{N_u=800}{\hspace{-5mm}$\scriptstyle \Entunten=800$}
\psfrag{Theorie}{\hspace{-5mm}\scriptsize theory}
\psfrag{Zeit}{$t[\frac{\hbar}{\Delta E}]$}
\includegraphics[width=7.5cm]{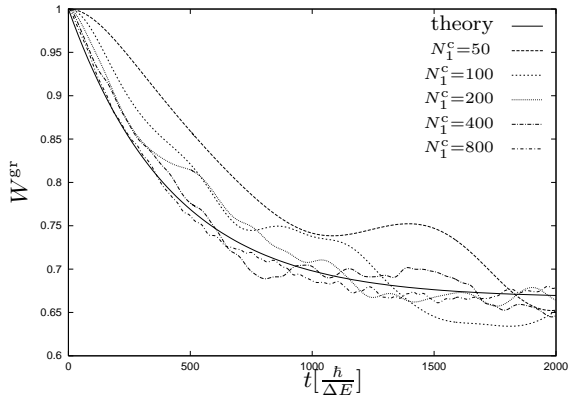}
\caption{Numerical simulation for the relaxation period. The predictions from the rate equation get better with increasing container system, $\Entartung{c}{1}$.}
\label{fig:ways:sec:num:1}
\end{figure}

The numerical results are displayed in \reffig{fig:ways:sec:num:1}. 
The solid line is the prediction from theory. 
Obviously the theoretical predictions are not accurate for \qmarks{\fewlevel} container environments. 
This is due to the fact that the replacement of actual quantities by their \hsav\ is a valid approximation for high dimensional {\hilbertspace}s only. 
Furthermore, for the \fewlevel\ cases the iteration step times that have to be longer than $\tau_1$ are rather long, because already $\tau_1$ is long. 
This means that the recursion cannot really be replaced by the differential equation in \refsec{chap:ways:sec:rateeq}.
This essentially shifts the corresponding curves to later times, compared to the theoretical prediction. 
All those effects vanish if the container system becomes sufficiently big. 
The simulation for $\Entunten=800$ is obviously in good agreement with the theoretical prediction.

\section{Summary and Conclusion}

We considered a two-level system coupled to an environment specified
only by two resonant energy bands. We showed that for systems of this
type (or rather any bi-partite quantum system) the entropy of the
considered system cannot increase without increasing
system-environment correlations, if the full system evolution is
unitary.

Further more we solved the Schr\"odinger equation for the full system
starting with an initial pure product state. Under some conditions
concerning interaction strength, band width and state density of the
environment-system we find a merely statistical energy transfer or
relaxation process, that may simply be described by some transition
rates. This result has been derived on the basis of a theory involving
Hilbert space averages and confirmed numerically.

We thus conclude that for statistical decay behavior of exited states
in quantum systems neither thermal nor infinite baths are necessary,
just as well as the factorizing condition seems neither tenable nor indispensable.


\end{document}